\documentclass[12pt]{amsart}

\usepackage{amssymb}
\usepackage{bm}
\usepackage{graphicx}
\usepackage{psfrag}
\usepackage{color}
\usepackage{enumerate}
\usepackage{enumitem}
\usepackage{url}

\usepackage{hyperref}
\definecolor{darkblue}{RGB}{0,0,160}
\hypersetup{colorlinks,citecolor=black,filecolor=black,%
            linkcolor=darkblue,urlcolor=darkblue}

\newcommand{\excise}[1]{}%{$\star$\textsc{#1}$\star$}

\theoremstyle{definition}

\newtheorem*{defn*}{Definition}

\newcommand{\ring}[1]{\ensuremath{\mathbb{#1}}}

\newcommand\RR{\ring{R}}

\newcommand\kk{\Bbbk}

\newcommand\ve{\varepsilon}

\newcommand\goesto{\rightsquigarrow}

\begin{document}%%%%%%%%%%%%%%%%%%%%%%%%%%%%%%%%%%%%%%%%%%%%%%%%%%%%%%%%
%%%%%%%%%%%%%%%%%%%%%%%%%%%%%%%%%%%%%%%%%%%%%%%%%%%%%%%%%%%%%%%%%%%%%%%%

\mbox{}\vspace{-4ex}
\title[Fruit flies and moduli]
	{Fruit flies and moduli: interactions\\
	between biology and mathematics}
\author{Ezra Miller}%\\\\(DRAFT---DO NOT DISTRIBUTE)
\address{Mathematics Department\\Duke University\\Durham, NC 27708}
\urladdr{\url{http://www.math.duke.edu/~ezra}}

\makeatletter
  \@namedef{subjclassname@2010}{\textup{2010} Mathematics Subject Classification}
\makeatother
\subjclass[2010]{Primary: ??X\#\#, ??X\#\# Secondary: ??X\#\#, ??X\#\#}
% Primary:
% 13C05 Structure, classification theorems (theory of modules and ideals)
% 05E40 Combinatorial aspects of commutative algebra
% 20M25 Semigroup rings, multiplicative semigroups of rings [See also 16S36, 16Y60]
% Secondary:
% 05E15 Combinatorial aspects of groups and algebras
% 13F99 Arithmetic rings and other special rings--none of the above but in this section
% 13A02 Graded rings [See also 16W50]
% 13P99 None of the above, but in this section (Computational aspects and applications)
% http://msc2010.org/MSC-2010-server.html
% Removed (from mesoprimary.tex):
% 20M14 Commutative semigroups
% 20M30 Representation of semigroups; actions of semigroups on sets
% 20M13 Arithmetic theory of monoids
% 14M25 Toric varieties, Newton polyhedra [See also 52B20]
% 68W30 Symbolic computation and algebraic computation


\date{31 July 2015}

% \begin{abstract}
% ...
% \end{abstract}

\maketitle

%%%%%%%%%%%%%%%%%%%%%%%%%%%%%%%%%%%%%%%%%%%%%%%%%%%%%%%%%%%%%%%%%%%%%%%%%
%section{Introduction}\label{s:intro}%%%%%%%%%%%%%%%%%%%%%%%%%%%%%%%%%%%%
%%%%%%%%%%%%%%%%%%%%%%%%%%%%%%%%%%%%%%%%%%%%%%%%%%%%%%%%%%%%%%%%%%%%%%%%%

%
\begin{excise}{%
  Biology has a long track record of interaction with applied
  mathematics and statistics, but direct communication with pure
  mathematics has been more rarified for biology than for other natural
  sciences such as physics or chemistry.  This state of affairs is
  % in the process of
  beginning to change (cf.\ M.~Reed's article in this issue), in part
  because certain topics within pure mathematics are
  % becoming more accepted as applied mathematics,
  widening their purview, and in part because mathematicians and
  biologists are finding more sophisticated ways to quantify
  biologically relevant
  % [phenomena; objects; structures; processes]
  structures and processes.  My purpose here is to demonstrate, by
  example,
  % the kinds of interplay
  % some
  types of exchanges between biology and pure mathematics that are
  possible, and likely to become increasingly common, in current and
  future research.

  0. Introduction:
     The article as a whole is about two-way conversations between math
     and biology, with an emphasis on the novel math.
     The subject of the talk is multigraded algebra of fruit fly wing
     topology.  The point is that multiparameter persistent homology is
     a multigraded module over a commutative polynomial ring.  Thinking
     statistically about such objects raises a lot of practical
     questions that turn out to be deep algebraic geometry and/or
     combinatorics.  Some of those deep problems can be treated with
     geometrically stratified statistics.
  
  I. Evolutionary origins of saltation (discrete topological
     changes in phenotype, such as adding a new segment or limb)
     Example for us, because this sort of thing can be tested,  
     tweaked, and selected for: fruit fly wing veins.
  
  II. Persistent homology as data summary, to see whether the    
      tweaking and selecting really has the suspected effect.
  
  III. Mathematical questions raised by statistical considerations
     in this context, one of them being how to do statistics when 
     sampling from algebraic varieties (moduli spaces).  Also
     perhaps some biological avenues suggested by the mathematical
     methods.
  
  IV. Geometric probability on stratified spaces, noting the
  
  V. Evolutionary genesis of topic IV.
}\end{excise}%

Possibilities for using geometry and topology to
% treat
analyze statistical problems in biology raise a host of novel
questions in geometry, probability, algebra, and combinatorics that
demonstrate the power of biology to influence the future of pure
mathematics.  This is a tour through some biological explorations and
their mathematical \mbox{ramifications}.

\subsection*{A biological hypothesis}%%%%%%%%%%%%%%%%%%%%%%%%%%%%%%%%%

Evolution sometimes results in discrete
% topological
morphological differences
% in phenotype from one population to an immediate
among populations that diverge from a common source.  This
``saltation'' can occur with features quantified by integers---limbs,
segments, petals, teeth, or digits (humans are occasionally born with
six fingers); or quantified by other discrete invariants, such as
tesselation patterns---seeds in flowers or protomers in virus capsids.
Biology has explanations of how populations that already exhibit a
varying trait can lead to populations in which one or the other
dominates.  The question here is:
% what generates the topological novelty in the first place.
what mechanism generates topological variation in sufficient quantity
for selection to act?

Take the fruit fly, for example.  The normal \textsl{Drosophila
melanogaster} wing depicted on the left differs from the abnormal
other two in topology as well as geometry.
% See
% http://www.sflorg.com/sciencenews/scn041906_02.html
% for images of different species' wings.
$$%
\includegraphics[height=23mm]{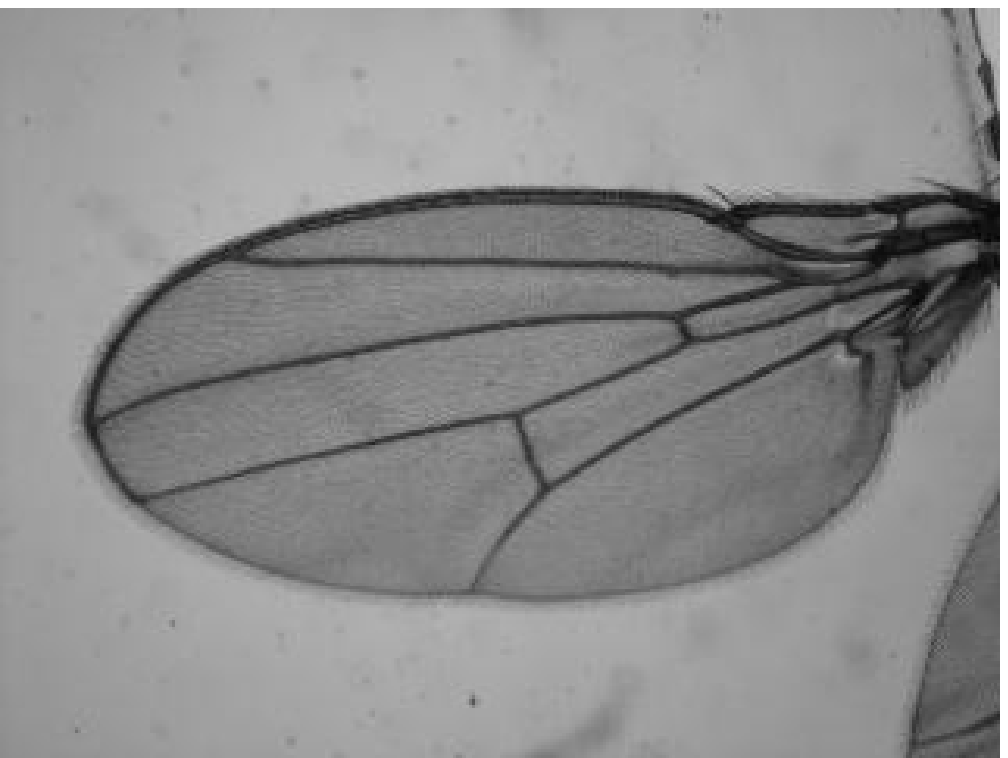}
\qquad
\includegraphics[height=23mm]{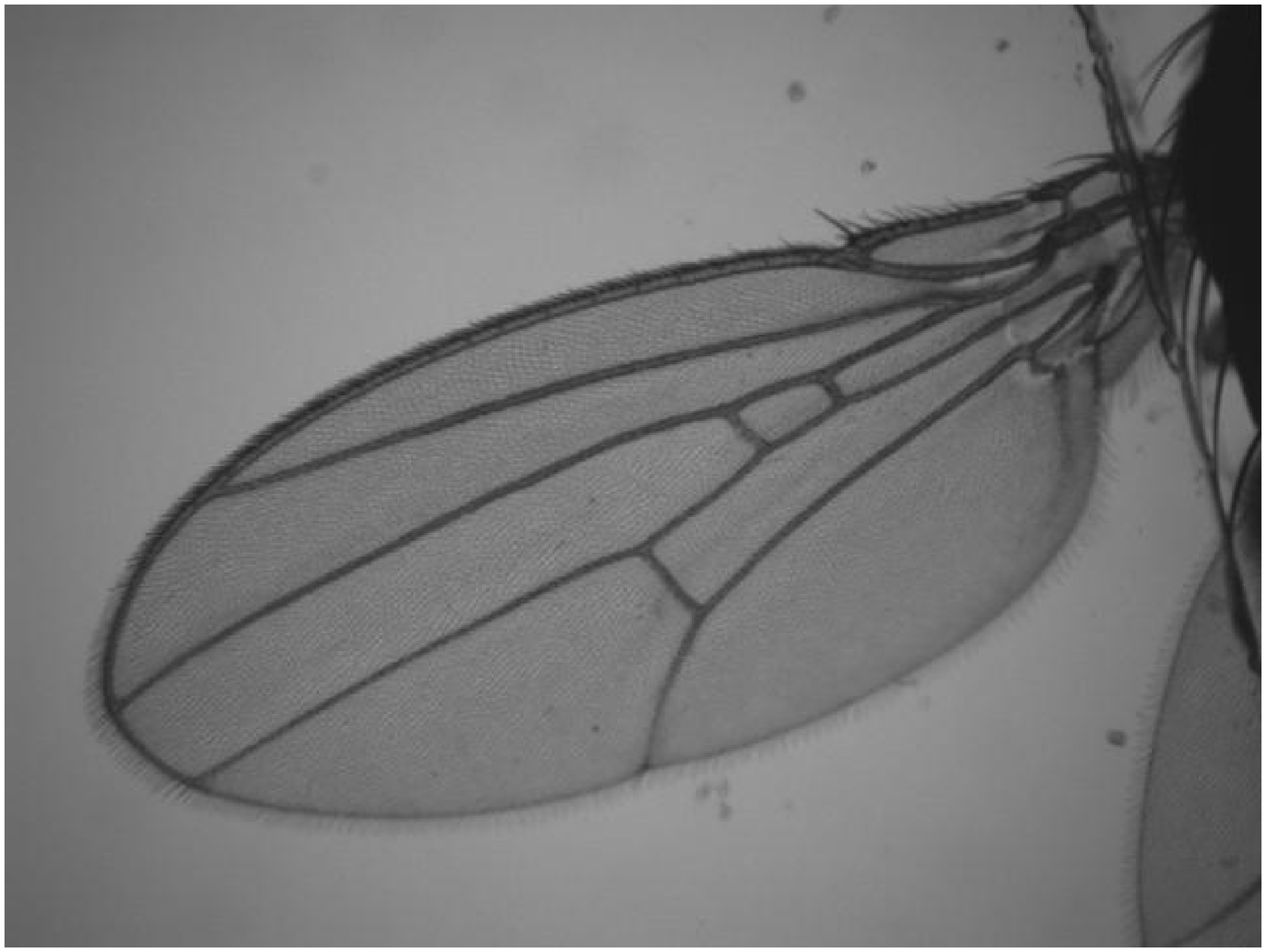}
\qquad
\includegraphics[height=23mm]{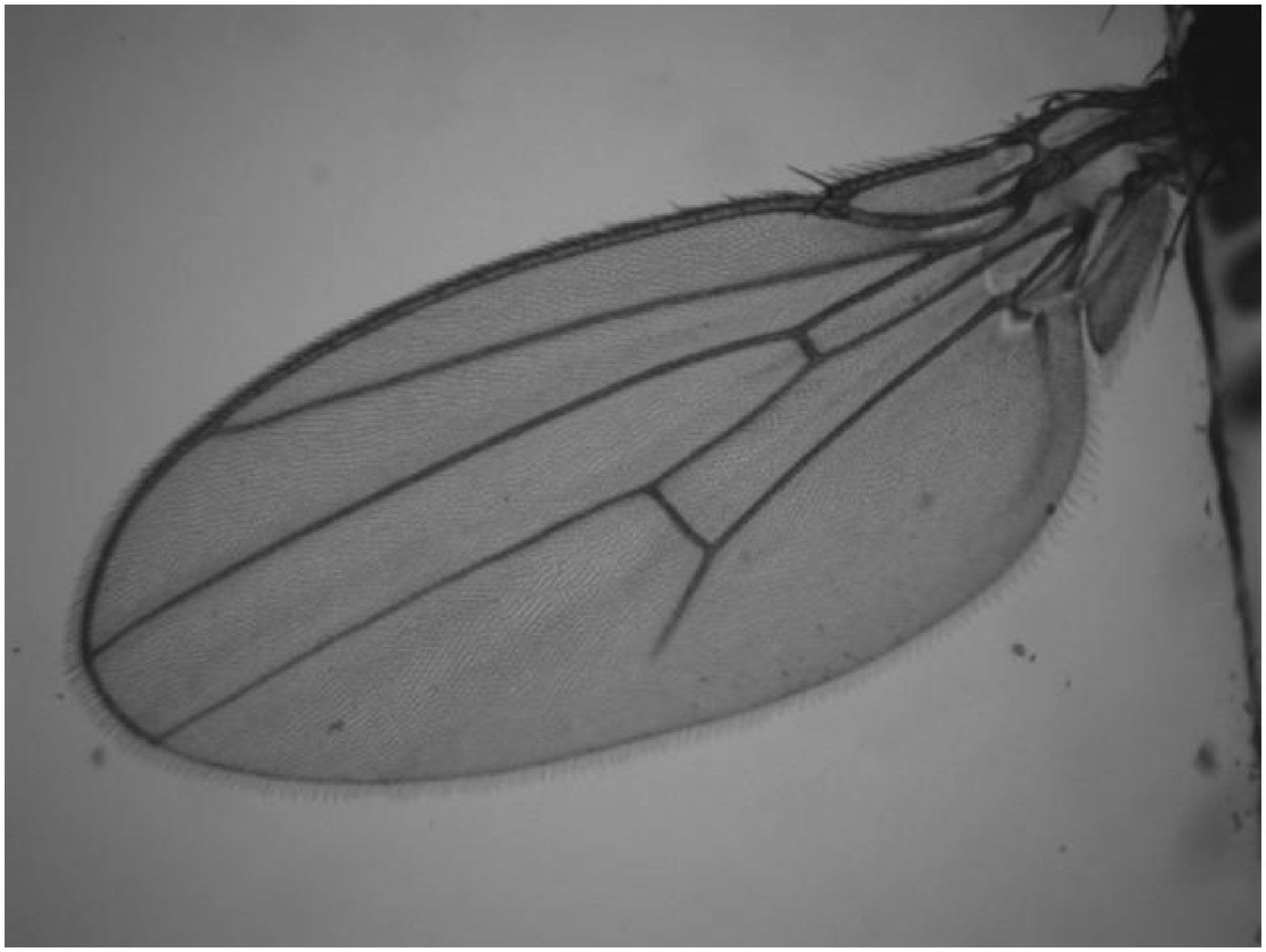}
$$
Indeed, mathematically the veins in each wing can be abstracted as an
embedded planar graph, with a location for each vertex and a contour
for each arc.  The graph in the middle has an extra edge, and hence
two extra vertices, while the graph on the right is lacking a vertex.
These topological variants, along with many others, occur in natural
\textsl{D.$\:$melanogaster} populations, but rarely.  On the other
hand, different species of \textsl{Drosophila} exhibit a range of wing
vein topologies.  How did that come to be?  Wing veins serve several
key purposes, as structural supports as well as conduits for airways,
% http://www.sdbonline.org/sites/fly/aimorph/trachia.htm#dafka
nerves, and blood cells, among other things \cite{blair07}.  Is it
possible that some force causes aberrant vein topologies to occur more
frequently than would otherwise be expected in a natural
population---frequently enough for evolutionary processes to act?

Results from biologist Kenneth Weber, and later with more power by
David Houle's lab,
% at Florida State
show that selecting for continuous wing deformations
% yields
results in skews toward
% increased prevalence of
% oddly shaped
deformed wings with normal vein topology
% as expected
\cite{weber90,weber92,houle03},
% But these selection experiments yield
but also---unexpectedly---much higher rates of topological novelty.
This latter claim, which is unpublished and has yet to be tested
statistically, suggests a fundamental biological hypothesis:
topological novelty arises at the extreme of selection for
continuous~shape~characteristics.
% To test the hypothesis, it would be ideal to plot

\subsection*{Wings to modules}%%%%%%%%%%%%%%%%%%%%%%%%%%%%%%%%%%%%%%%%

% It might be
% It is potentially possible to test this hypothesis
This hypothesis could potentially be tested using persistent homology,
a tool for data analysis that uses computational topology to assign
modules over polynomial rings to subsets $X \subseteq \RR^n$
\cite{multiparamPH}.
% % but
% % and
% That possibility raises a host of novel questions in algebra,
% geometry, and combinatorics that demonstrate the power of biology to
% influence the future of pure mathematics.
% % must be answered first
% % stand in the way.
% % The suitability of
% The multiparameter version of persistent homology intended here
This tool is a good candidate because of its ability to emphasize
differences in stratification among otherwise similar
subsets~of~$\RR^n$.

% The general theory would take too long to present in this brief
% note, but for an embedded graph $X \subseteq \RR^2$ it's not hard.
Take our case an of embedded graph $X \subseteq \RR^2$, for example.
For any nonnegative real numbers~$r$ and~$s$, let~$X_r^s \subseteq X$
be the set of points at distance at least~$r$ from every vertex and
within~$s$ of some edge.  Thus $X_r^s$ is obtained by taking the union
of the balls of radius~$r$ around the vertices away from the union of
$s$-neighborhoods of the edges.  In the following magnified portion of
the middle wing, $r$
% (in blue)
is approximately twice~$s$:
% (in red):
$$%
\includegraphics[height=23mm]{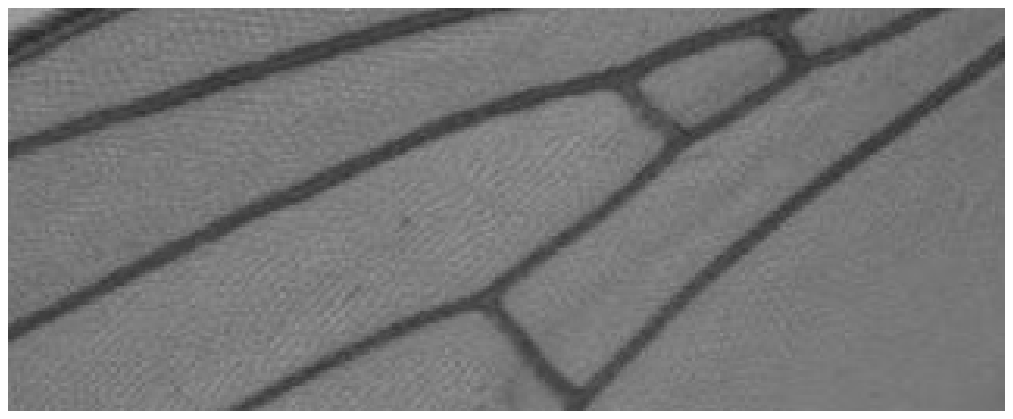}
\quad
\raisebox{10mm}{$\goesto$}
\quad
\includegraphics[height=23mm]{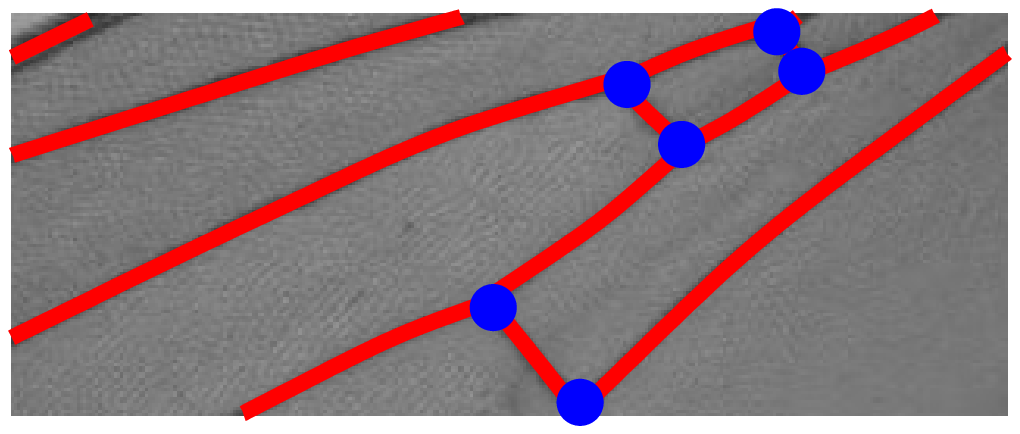}
$$
The homology $H_i(X_r^s)$ with coefficients in a field~$\kk$
% (think $\kk = \ZZ/2\ZZ$ for utmost simplicity)....
counts connected components or loops of $X_r^s$, when $i = 0$ or~$1$,
respectively.  Introducing a new vertex to an edge of~$X$ tends to
create connected components and destroy cycles when $r \gg s$, because
the balls around vertices protect them from the expanding edges.  The
precise relations between $r$ and~$s$ that alter the topology
of~$X_r^s$ depend on the geometry of~$X$, such as the angles between
edges at a given vertex, and that is the point: persistent homology
uses topological invariants as measures of geometry.
% \comment{The form of multiparameter persistent homology described
% here can be viewed as a generalization of persistent intersection
% homology in which interactions among strata are tuned by relations
% between the parameters.  Why might I say this?  Because this
% generalization was genuinely inspired by the fly wing phenomics
% project.  I would in fact write the first sentence of this comment
% in the article if there was a source to cite (beyond the
% Bendich-Harer definition of PIH) concerning the relation between
% this multiparameter persistence and PIH.}

To keep the data structure finite, the parameters~$r$ and~$s$ can,
without loss of significant information, be restricted to integer
multiples of a small positive length~$\ve$.  The
% module $M_i(X)$ is the
persistent homology of the graph~$X$ with the two parameters $r$
and~$s$ is then defined to be the direct sum $M_i(X) = \bigoplus_{r,s}
H_i(X_r^s)$ of all of the homology groups.  It is a bigraded module
over the polynomial ring $\kk[x,y]$: the variables $x$ and~$y$ act
on~$M_i(X)$ by comparing the homology of $X_r^s$ with that of
$X_{r-\ve}^s$ and $X_r^{s+\ve}$, respectively.

%\subsection*{Persistent homology as data summary}%%%%%%%%%%%%%%%%%%%%%

% ...to see whether the tweaking and selecting really has the suspected
% effect.
% 
% - dataset = point cloud; persistent homology tells you something
%   about the topology or geometry of the dataset.
% 
% - data object = geometric thingy; dataset = bunch of geometric
%   thingies; persistent homology tells you something about each data
%   point and thereby reduces dataset to bunch of persistence diagrams.

% Beyond the practicalities of statistical application, there lies the
% question of whether these---or other---discrete proxies for
% continuous moduli admit transparent topological or geometric
% interpretation.

Persistent homology with only one parameter
\cite{robins99,frosini-landi01}, instead of two or more, results in a
module over a polynomial ring in one variable.  This case is much
better studied, in part because it behaves more tamely.  In
particular, there is a finite, computable set of topological
features---connected components,
% when $i = 0$
% or
loops
% when $i = 1$
or, in the general case, features of higher dimension---each of which
has well defined parameters where its ``birth'' and ``death'' occur,
such that every homology class is a direct sum of these
features~\cite{edel-letc-zomo02}.  For fly wings, or arbitrary
multiparameter situations, where the homology groups
% indexed by parameters
record the topology of several
% filtrations
increasing chains of subsets of a single topological space, no such
clean description is possible \cite{multiparamPH}.  However,
alternative presentations of modules from combinatorial commutative
algebra \cite{irredRes} (or see~\cite[Chapter~11]{cca}), based
essentially on the theory of primary decomposition, can be understood
topologically in terms of birth and death parameters.  Such
understanding is necessary if statistics on sets of bigraded modules
are to be interpretable biologically.

\subsection*{Moduli as statistical sample spaces}%%%%%%%%%%%%%%%%%%%%%

Persistent homology summarizes a sample of fly wings by transforming
it into a sample of modules.  It is a general principle of statistics
that to analyze samples from a set of objects one needs sufficient
understanding of the set of all objects from metric, probabilistic,
and sometimes combinatorial perspectives.
% distance between pairs of objects and distributions on the set of
% objects.
How far apart are pairs of objects?  How likely is each object to be
selected at random?  Mathematics excels at placing coherent structures
on sets of all objects of a given type.  The resulting ``moduli
spaces'' pervade geometry of many sorts---differential, algebraic,
arithmetic, complex, discrete---and also theoretical physics and
topology, though in the latter field they are called classifying
spaces.
% Despite being ubiquitous and in some cases our having substantial
% understanding of their geometry and combinatorics,
But despite their ubiquity and in some cases our substantial
understanding of their geometry and combinatorics, less is known about
the probability and statistics of sampling from~them.

Like many moduli spaces, the ones parametrizing bigraded modules over
$\kk[x,y]$ are quotients of algebraic varieties by continuous group
actions \cite{multiparamPH}.  This makes the moduli spaces complicated
% singular non-Hausdorff
unions of manifolds of varying dimension.  One possibility, covered in
the next section, is to develop geometric methods to analyze samples
from such ``stratified spaces''.  Another, which tends to be favored
a~priori for computational reasons, is to use discrete invariants as
proxies for the continuous moduli.  For bigraded modules, these
discrete invariants include
\begin{itemize}
\item%
single-parameter persistence by tracing zigzags through the
% bigraded components
groups~$H_i(X_r^s)$ \cite{zigzagPH};
\item%
Hilbert series, meaning the dimensions of the vector spaces
$H_i(X_r^s)$, disregarding all of the homomorphisms between them;
\item%
rank invariants, which take into account the ranks of the
homomorphisms but not their precise algebraic structure
\cite{multiparamPH,cag-dif-fer10};
\item%
% multigraded
Betti numbers, which record discrete homological invariants of the
module \cite{knudson08}.
\end{itemize}
% % The suggestion Immediately it becomes necessary to know
% % \comment{that's a long phrase; tighten} whether
% % The following questions are, given the fly wing motivation,
% % practical ones
% The potential for statistical applications leads to several new
% sorts of questions for moduli spaces.  They are pragmatic for
% bigraded modules over $\kk[x,y]$, given the fly wing context, but
% they make sense for any moduli space---or indeed \comment{verify;
% modify as necessary}, any stratified space whatsoever.  The salient
% [look up...] observation is that
Any discrete invariant subdivides the moduli space into regions where
the invariant is constant.  Understanding the nature of these
subdivisions is a pragmatic matter for bigraded $\kk[x,y]$-modules,
given the fly wing context, but (modifications of) some of these new
sorts of questions make sense for any moduli space, or indeed any
stratified~space.
% whatsoever.
% with discrete invariants.
\begin{enumerate}
\item%
What metric or combinatorial properties do these subdivisions possess?
For example, do the regions have equal dimension and roughly equal
size, or
% is there a small set of
are there a few big regions (or only one) of top dimension and bunch
of smaller~ones?
\item%
% In either case,
What distribution of discrete invariants are expected from a given
(biological) problem?  Might the discrete invariants be expected to
distinguish between the
% quiver reps actually
modules produced by
% honest
applied situations even if the regions aren't of similar size?
\item%
General geometric statistical question: what (natural) measures should
be placed on a set of discrete invariants, given the geometry of the
moduli spaces?
\item%
Can the continuous variation be captured
% ``numerically''
discretely
% or \emph{effectively}:
to desired precision?  More precisely, is there a family (indexed
by~$n$) of sets of discrete invariants such that letting $n \to
\infty$ results in an increasingly fine subdivision?
% Orally: Figure out whether (say) paths of length bounded by n
% through the quiver work this way.  Note that we're not talking
% merely about zigzag paths, which are weakly monotone in each
% coordinate, but rather all paths through the quiver, including those
% that don't induce equioriented paths.
\end{enumerate}

% Example: Betti numbers are (Zariski!) upper semicontinuous $\implies$
% expect ``generic'' behavior if the modules in question are generated
% suitably randomly; does it actually happen this way in biological
% problems?
% % Orally: In the fly wing analysis, it's not so much variation of
% % Betti numbers within a single connected (or irreducible) component
% % of the moduli space, but instead variation of Hilbert function.
% % Remember that the grading is given by scale, or equivalently by
% % distance.  When vein strata move around, that alters the location --
% % in the plane of the wing -- where the topological features are born
% % or die.  New strata create additional generators and/or relations in
% % the corresponding multigraded modules.  Both of these kinds of
% % variation change the Hilbert function and therefore the multigraded
% % Hilbert scheme and therefore (kal v'chomer) the connected component
% % of the moduli space in question.  Specific question to ask in this
% % case: do the multigraded modules from fly wing multiparameter
% % persistence have Betti numbers that are generic for their
% % multigraded Hilbert scheme?

\subsection*{Geometric probability on stratified spaces}%%%%%%%%%%%%%%

% \mbox{}
% 
% \noindent
% \comment{go slower here}

% [As the reader proceeds through this section, they might have no
% idea that it was motivated by biology, and they might deem it pretty
% flimsy to claim the fly wing project as sufficient motivation.  Then
% % shock
% surprise them with the actual biological motivation (next section,
% on phylogenetic trees)]

As we have seen for fly wings, statistical problems where the sample
objects are more complicated than vectors in vector spaces naturally
% land in the realm of geometry
% % % combinatorics or (often algebraic)
% % Furthermore, the questions that one asks in those realms often feel
% % foreign (to mathematicians as much as the statisticians, biologists,
% % or anybody else in the picture).
lead to sampling from stratified spaces.
% Whenever you have to deal with the collection of all objects of a
% certain type, you're dealing with a moduli space.  This isn't news
% to, say, theoretical physicists, who already have theories
% explicitly referring to moduli spaces, such as moduli spaces of
% vacua, or of various sorts of branes, or more general instantons [be
% sure to get right the relation between branes and instantons].  But
% for biologists (or statisticians), the concept of moduli space feels
% far removed, possibly because the objects of interest to its
% practitioners defy precise definition.  (I was once told by a
% ``mathematical biologist'' that a species is a DNA strand; such a
% reduction is antithetical to the biological species concept.)
The goal of geometric statistics in this setting is, like in ordinary
linear statistics, to
% % get a handle on
% % describe
% % understand
% quantify variation, including how much
% % is present and in what directions it occurs.
% of it occurs in which directions within the sample space.
identify, describe, summarize, or make inferences about an unknown
probability distribution on the sample space from which the sample
points are assumed to be drawn.  To that end, it is crucial to
understand the opposite problem, from probability theory: given a
distribution on the relevant sample space, how do samples from that
distribution behave?

% Intuitively, samples of increasingly large cardinality
% % should
% provide increasingly accurate information about the distribution.  To
% understand some of the challenges of geometric statistics, consider
% % first the ordinary linear case.
% the simplest summary of a distribution: a
% % point
% \emph{mean} about which it is centered.
% % skip: line or higher dimension linear subspace summary = PCA;
% Laws of large numbers assert that 
The simplest summary of a distribution is a point---an average or
\emph{population mean}---about which the distribution is centered.
% ``sample mean'', to distinguish it from the ``population mean''
% (``theoretical mean'') ... finite sample
Laws of large numbers assert that means of increasingly large random
samples from a distribution converge to a population mean.  Statistics
requires knowledge of the expected difference between a sample mean
and population mean.  Central limit theorems help quantify that
difference by describing the variation of sample means around
population means.
% asymptotics of convergence in CLT required for estimates of error,
% such as confidence regions
In ordinary statistics, for example, when the sample space is the real
line, the central limit theorem dictates that sample means vary around
the population mean according to a distribution that is, in the limit
of infinite sample size, Gaussian.
% after appropriate rescaling to account for increasing sample size

% \comment{progression: linear, smooth, stratified}
% 
% \comment{progression: linear, linearly approximable, fully geometric}

In Euclidean space, basic concepts such as mean, expectation, and
average coincide and therefore admit multiple equivalent
characterizations, such as via least squares or arithmetic average.
Already thinking about asymptotics of samples from smooth
manifolds---let alone singular spaces such as the moduli spaces
relevant to fly wings---requires a radical shift in perspective, as
compared with samples from linear spaces, because different
characterizations lift to different notions in curved
settings~\cite{huck12}.  Moreover, for many of these notions, such as
Fr\'echet mean defined by least squares, the minimizer is not unique:
what is the average of the north pole and south pole on the sphere?
It is the
% whole
entire equator.  (This explains the phrase ``a population mean'' in
the previous paragraph, as opposed to ``the population mean''.)
Nonetheless, laws of large numbers hold \cite{ziezold77,bp03}, and
central limit theorems exist in various situations
\cite{jupp88,hl96,hl98,bp05,huck11}, such as when the data are
concentrated near a Fr\'echet mean.  (Many of these theorems were
motivated by biology; read the title of \cite{huck11}, for instance.)
For additional background and references concerning statistics on
manifolds,
% especially including details about numerous concept of mean
see~\cite{huck12}.

Statistics on smooth manifolds relies on approximation of the manifold
by its tangent space, which is Euclidean.  Once a metric on the
manifold has been specified---a necessary and often nontrivial step
for statistics, because of the need to know how far apart sample
objects are---the exponential map at a point~$x$ takes a
neighborhood~$U$ of~$0$ in the tangent space~$T_x$ to a neighborhood
$\exp(U)$ around~$x$.  Ordinary probability in the vector space~$T_x$
is transformed into geometric probability on the smooth sample space
via the exponential map at~$x$, which is close to an isometry when~$U$
is small.
% The rescaling in CLTs eventually takes you into a small neighborhood
% anyway...
In particular, central limit theorems on smooth manifolds can be
interpreted as describing variation of sample means around a
population mean by
% constructing limiting distributions on sample spaces by
% % pulling back
pushing forward the linear
% situation
setup along an exponential~map.

In the singular
% (that is, non-smooth)
setting of stratified spaces, there is no
% possibility
general method to compare with or reduce to ordinary linear
probability and statistics in the tangent space at
% the mean (or, with nonuniqueness, a~mean).
a mean.
% ...ad hoc...
% Progress so far includes laws of large numbers and central limit
% theorems in specific simple examples,
The types of sample spaces~$M$ intended here are those possessing a
\emph{topological stratification} (see \cite{gm88} or
\cite{pflaum01}): an expression as a disjoint union $M = M_0 \cup M_1
\cup \cdots \cup M_r$ of \emph{strata} such that for all~$d \in
\{0,\ldots,r\}$,
\begin{itemize}
\item%
the stratum $M_d$ is a manifold,
\item%
$M_0 \cup \cdots \cup M_d$ is closed in~$M$, and
\item%
for every pair $x,y \in M_d$ there is a homeomorphism $M \to M$ that
takes $x$ to~$y$ and takes each stratum to itself.
\end{itemize}
The third condition ensures that the topology of~$M$ and its
stratification behaves precisely the same way near~$x$ as it does
near~$y$.  Examples include graphs (or networks), whose strata are
vertices and edges; polytopes, whose strata are (relatively open)
faces; and real (semi)algebraic varieties, whose strata consist of
classes of singular points.  The tangent space~$T_x$ to a stratified
space~$M$ at a point~$x$ is a cone over a stratified space of
dimension one less than~$\dim(M)$.  If $M$ is already a cone with
apex~$x$, then $T_x \cong M$ is as complicated as $M$ itself;
capturing the local structure of a stratified space near a point need
not simplify the geometry the way it does in the smooth setting.

Cases where stratified laws of large numbers and central limit
theorems are known occur in specific simple examples where comparison
with linear spaces is possible, such as open books \cite{stickyCLT}
(unions of Euclidean half-spaces glued along their boundary
subspaces), isolated planar hyperbolic singularities (cones where the
singular point has angle sum~$> 2\pi$ instead of the ice-cream case
of~$< 2\pi$) \cite{dim2sticky}, and binary trees \cite{basrak10}.
% \comment{and weird spaces, like circles with stuff sticking out of
% it, constructed specifically to make stickiness subtle.  But I can't
% find any results by Stephan written down about this.}
But in general, de~novo geometric constructions are required.  Such
has been the avenue for the good deal of probability theory, including
laws of large numbers, that has been established in the generality of
nonpositively curved spaces (see~\cite{sturm03}), which are defined as
spaces whose triangles with given edge lengths are thinner than would
be expected from Euclidean geometry (see~\cite{bridson-haefliger}).
The metric structures of nonpositively curved spaces
% possess
induce a number of simplifying consequences, such as uniqueness of
Fr\'echet means, which have played important roles in the progress
thus far in geometric probability and statistics on stratified spaces.
Nonetheless, the promising interactions of nonpositive curvature with
geometrically stratified probability and statistics remain~in their
infancy.

\subsection*{Sticky means}%%%%%%%%%%%%%%%%%%%%%%%%%%%%%%%%%%%%%%%%%%%%

% Any general comparison with or reduction to ordinary linear
% probability and statistics would be complicated, at best:
The
% difficulty and
novelty of attempting statistics on stratified sample spaces is
exemplified by
% resolutely
nonclassical ``sticky'' phenomena that can occur at singularities.  In
Euclidean statistics, the mean of a finite set of points moves
slightly in any desired direction by perturbing the points.  This
intuition extends to manifolds, by linear approximation \cite{jupp88,
hl96, bp05, huck11}, but it can fail even in the simplest singular
sample spaces.  Consider the tripod, for instance, depicted at left:
$$%
% \begin{array}{@{}*4{c@{\qquad\qquad}}@{}}
% \psfrag{0}{\small$0$}
% \includegraphics[height=20mm]{spider.eps}
% &
% \psfrag{0}{\small$\,\mu$}
% \includegraphics[height=20mm]{delta.eps}
% &
% \psfrag{0}{\small$\,\mu$}
% \includegraphics[height=20mm]{wiggle.eps}
% &
% \psfrag{0}{\small$\,\mu$}
% \includegraphics[height=22mm]{nudge.eps}
% \end{array}
\psfrag{0}{\small$0$}
\includegraphics[height=20mm]{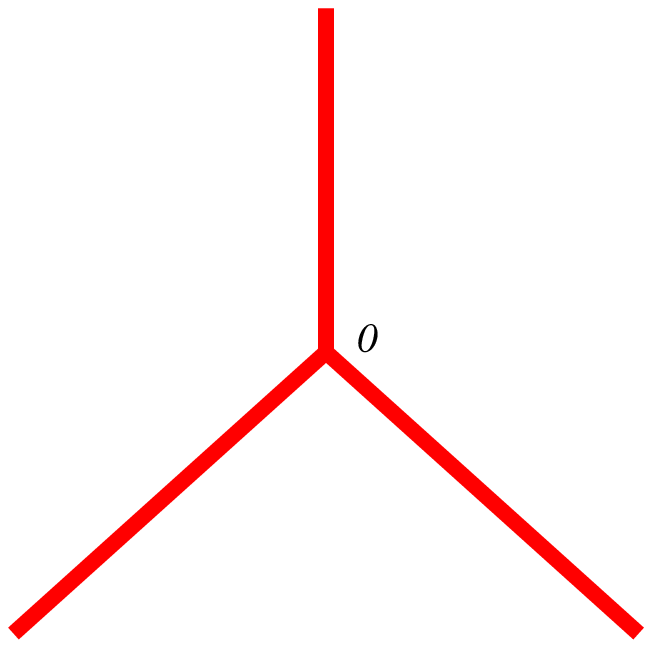}
\qquad\qquad
\psfrag{0}{\small$\,\mu$}
\includegraphics[height=20mm]{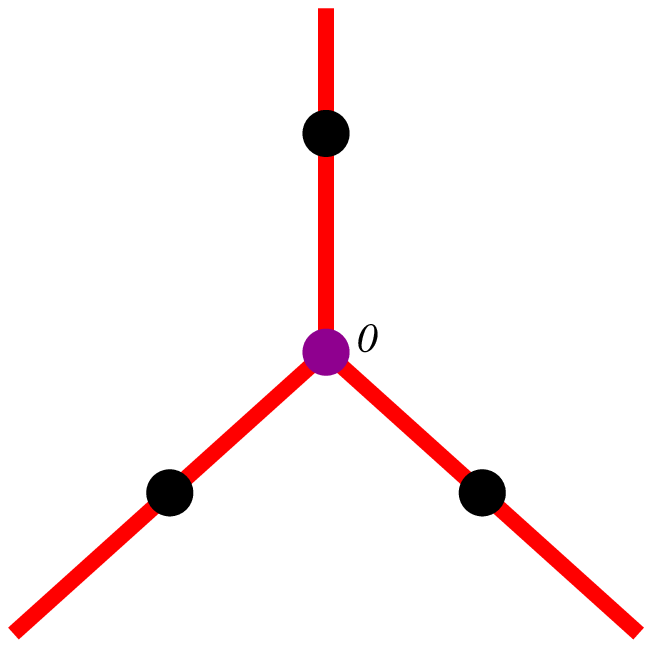}
\qquad\qquad
\includegraphics[height=20mm]{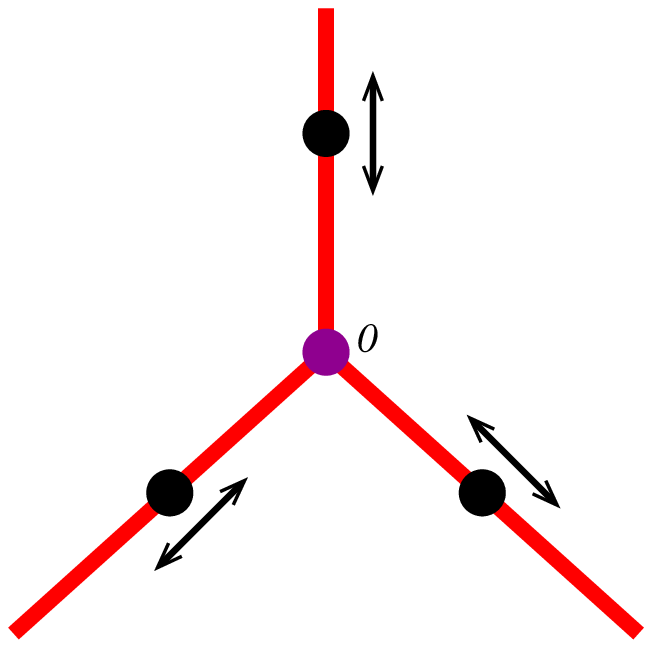}
\qquad\qquad
\raisebox{-1.6mm}{\includegraphics[height=21.6mm]{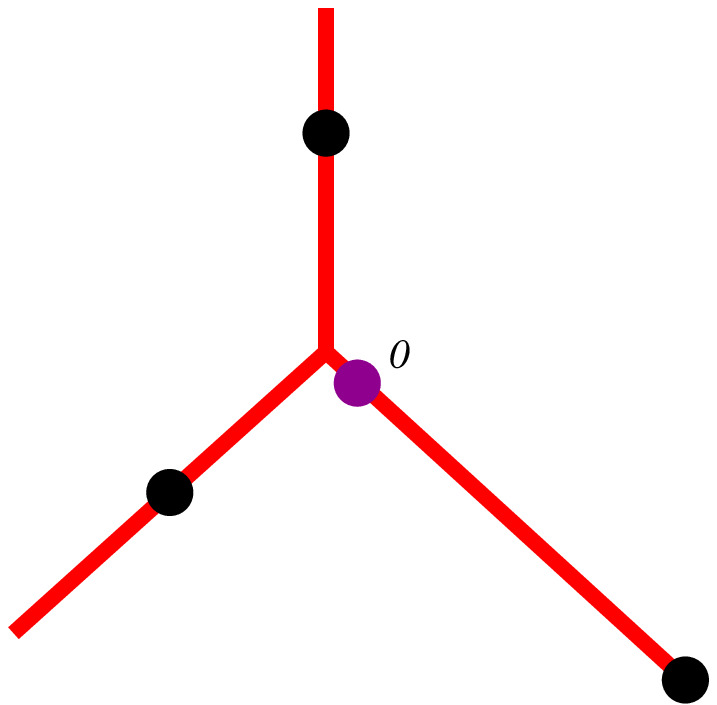}}
$$
In the center-left figure, the Fr\'echet mean~$\mu$ of the three
points on the legs is the origin~$0$, by symmetry.  But wiggling the
three points, as in the center-right figure, does not move the
Fr\'echet mean at all; one of the points would have to be moved more
than twice as far from the center, as in the final figure, to nudge
the mean onto its leg.

An open book with three pages is a product of tripod with vector
space~$\RR^d$.  (To get an arbitrary number of pages, replace the
tripod with a graph having an arbitrary number of rays emanating from
the center point.  It bears mentioning that every topologically
stratified space~$M$ is locally homeomorphic to an open book near any
point on a stratum of dimension $\dim(M) - 1$.  In other words, the
tangent cones to points on codimension~$1$ strata are open books.
Hence this example is universal in some sense.)  In an open book, the
mean sticks to the spine---the copy of~$\RR^d$ that is contained in
all three pages---when three similarly situated points are wiggled,
although that wiggling can move the mean in arbitrary directions
inside of the spine \cite{stickyCLT}.

With these examples and others in mind, a formal definition has been
developed~\cite{dim2sticky}:
% \begin{defn*}
let $\mathcal{M}$ be a set of measures on a metric space~$K$.  Assume
$\mathcal{M}$ has a given topology.  A \emph{mean} is a continuous
assignment $\mathcal{M} \to \{$closed subsets of~$K\}$.  A
measure~$\mu$ \emph{sticks} to a closed subset $C \subseteq K$ if
every neighborhood of~$\mu$ in~$\mathcal{M}$ contains a nonempty open
subset consisting of measures whose mean sets are contained in~$C$.
% \end{defn*}
% In terms of geometric probability,

Stickiness implies that it is possible for the means of large samples
from a distribution on a stratified space to lie in a subset of low
dimension, with positive probability (equal to~$1$ in some cases, such
as the tripod), even if the distribution being sampled is well behaved
\cite{blo13,basrak10,stickyCLT,dim2sticky}.  This contrasts with usual
laws of large numbers, where sample means approach the population mean
but almost surely never land on it---or on any given subset of low
dimension containing~it.  Thinking in terms of central limit theorems,
whereas in usual cases the limiting distributions have full support,
in sticky cases the limiting distributions can have components
supported on low-dimensional subsets of the sample~space.

Examples aside, central limit theorems on stratified spaces of any
generality have yet to be formulated, let alone proved.  Subtle and
deep behavior associated with the boundary between sticky and
non-sticky are still being discovered.  In particular, the distinction
between positive and negative curvature seems to be critical for
stickiness.
% To what extent is stickiness tied to notions of negative (sectional)
% curvature?  ...that is, means of distributions stick to a subset if
% the curvature is negative in the orthogonal direction...
% ...
% Get to the state of things: sticky with negative curvature, smeary
% with positive curvature.  Take from intro to Stephan's paper.
% ...
One common type of positive curvature, particularly in statistical
problems, appears when a flat or positively curved smooth manifold is
quotiented by
% an appropriate continuous
a proper Lie group action.  Shape spaces
(see~\cite{kendall84,shapeTheory})---including those
% of the sort
applicable to fly wings with constant topology
% ---one keeps track of the locations of the vertices---
by keeping track of the locations of the vertices of the graph---have
this form, for instance, being quotients of matrix spaces by actions
of
% groups of
rotations, scaling, or projective transformations.
% groups.
Huckemann \cite{huck12}
% and others
has shown essentially that when sampling from a stratified space that
is a quotient of this form,
% positively curved stratified space that is a quotient of a flat or
% positively curved implies
Fr\'echet means run away from singularities and hence land almost
surely in the smooth locus.  On the other hand, every case exhibiting
stickiness has negative curvature (in the sense of Alexandrov:
curvature bounded strictly above by~$0$;
see~\cite{bridson-haefliger}).  It remains
% a major
an open problem to formulate a condition, in terms of something like
negative sectional curvature, that allows means to run toward
singularities of the space for appropriate types of
distributions~on~it.

% ...
% 
% A theory of confidence regions is beginning to take
% shape~\cite{hotz-le14}: how large must a sample be to ensure with
% sufficient probability that the mean (set) lies in a given stratum?
% 
% ...

% \comment{this is more about the smooth case:} ...placement of data
% with respect to the cut locus of the mean set---that is, the points
% that have more than one shortest path to the mean set; mass near the
% cut locus pulls the mean in multiple directions at once.

% \comment{this is interesting, and probably good for the NSF proposal
% as motivation, but too vague for inclusion here}
% analogue with algebraic geometry:
% What can be said about the geometry or topology of a space when the
% only information given is a finite set of points in the space?
% ...finite (generic) samples from algebraic varieties provide enough
% information to compute their cohomology
% \cite[Proposition~1.3]{mustata1998}
% ... ...
% An overarching goal in this area:
% general analogy: probability in metric geometry acts as a stand-in for
% homology in algebraic geometry: H(point clouds in alg geom)
% $\leftrightarrow$ geometry of CLT for point clouds in metric geom
% ...Riemannian manifolds are much less rigid than algebraic varieties,
% so it is only reasonable to expect cohomological conclusions to come
% from asymptotic behavior of large samples.

Further potential implications for pure mathematics
% become clearer
emerge when considering how to recover curvature invariants from
asymptotics instead of the other way around.  In the smooth case, for
local samples---that is, sufficiently nearby the mean---accounting for
% (``modding out by'')
curvature reduces central limit theorems to the Euclidean version.
Thus, if the curvature is unknown but properties of the distribution
are known, then the curvature ought to be recoverable.  In singular
settings, attempting such a recovery could give rise to singular
analogues of smooth Riemannian curvature invariants.

\subsection*{Phylogenetic trees}%%%%%%%%%%%%%%%%%%%%%%%%%%%%%%%%%%%%%%

Moduli spaces of fly wing modules constitute just one of
% endless
myriad ways that geometric probability on stratified spaces can
arise.
% and fly wings were not the original motivation for the study of
% stratified statistics.
% And indeed,
Among those, nothing is special about biology,
% in this regard,
% although
except perhaps that its diversity of forms and the nature of their
variation lend themselves to geometric data analysis of this sort.
That said, the
% initial motivation for
genesis of stratified statistics came directly from another
% aspect of
evolutionary biology moduli~space.

One of the principal aims of systematic and evolutionary biology is to
determine relationships between species.  Trees representing these
kinships are reconstructed from biological data such as DNA sequences
or morphology.
% The reconstruction process
Experimental procedures generate distributions on the set of
phylogenetic trees in multiple ways.
% many gene trees, one species tree; one set of species, one dataset
% relating them, posterior distribution from probabilistic tree
% reconstruction algorithm \cite{rannala-yang97,mrBayes01}
For example,
% analyzing
the evolutionary history of
% DNA sequence of
a single gene across multiple individuals is represented by a ``gene
tree''.  Natural processes such as incomplete lineage sorting cause
gene trees sampled from a
% given set of species to differ in topology from one another and
% % possibly
% from the evolutionary history of the species themselves,
% % the history of population bifurcations leading to divergence.
% % ``species tree''
% even when the DNA sequences are sampled from the same set of
% individuals
set of individuals to differ in topology from one another and from the
evolutionary history of their set of species---the history of
population bifurcations leading to divergence
% ``species tree''
(see \cite{maddison97}, for example).
% Furthermore, the most likely gene tree topology need not agree with
% the species tree topology \cite{degnan-rosenberg06}.
% However, species trees are usually reconstructed from gene trees,
% and a major open question is how best to accomplish~this.
Another crucial example occurs not from the data but from the method
of inference:
% one set of species, one dataset relating them,
the problem of phylogenetic tree reconstruction is intractable enough
that probabilistic methods are commonly used, resulting in posterior
distributions
% even in the context of one set of species and one dataset relating
% them
instead of a single optimum \cite{rannala-yang97,mrBayes01}.

Mathematically speaking, a phylogenetic tree on a given set of species
is a connected metric graph, with no loops, whose vertices of
degree~$1$ (``leaves'') are labeled by the species.  The introduction
by Billera, Holmes, and Vogtmann of an appropriate moduli space for
the problem, namely the space of phylogenetic trees \cite{bhv01},
initiated a
% flurry
surge of activity attempting to mine the combinatorics and geometry of
the space to devise statistical methods.  And the combinatorics is
formidable: for $n$ species, the tree space $\mathcal{T}_n$ is a
polyhedral stratified space composed of $(2n-3)!!$ Euclidean orthants
of dimension~$n-2$.
% Nonetheless,
Despite its complexity, $\mathcal{T}_n$
% yielded
% acquiesced:
succumbed: the advent of a
% fast
polynomial time algorithm for shortest paths in tree space
\cite{owen-provan10} made it possible to compute Fr\'echet means
in~$\mathcal{T}_n$ \cite{bacak12,centroids} based on probability
theory for nonpositively curved spaces \cite{sturm03}.

The geometric probability on stratified spaces in the previous two
sections was initially developed specifically to understand the
behavior of Fr\'echet means in the moduli space~$\mathcal{T}_n$.
% of phylogenetic trees on $n$ species.
The simple examples \cite{stickyCLT, dim2sticky} deal with informative
local subsets of~$\mathcal{T}_n$.  In addition to those, efforts are
underway to prove laws of large numbers and central limit theorems in
the global context of~$\mathcal{T}_n$ itself \cite{blo13,blo14}.

Stickiness in tree space has a concrete, meaningful biological
interpretation (although the jury is out on the extent to which the
interpretation reflects reality).  Points in strata of lower dimension
represent phylogenetic trees with one or more non-binary vertices.
Biologically, these are unresolved phylogenies: one species diverges
simultaneously into three
% or more
new species, for example, instead of first splitting into two new
species followed by another binary divergence event from one of the
two.  Sets of phylogenetic trees from biological experiments often
contain evidence for many or all
% three
of the possible sequences of binary divergence events that resolve a
given
% ternary
multiple divergence.  Stickiness implies that the mean tree will
contain
% the unresolved ternary
an unresolved divergence
% unless one of the three sequences occurs
whenever there is insufficient strength of pull toward any one of the
% three options
resolving binary sequences to support the conclusion it represents.
The picture to keep in mind is the tripod, whose stickiness we saw
earlier: it is
% actually
the tree space~$\mathcal{T}_3$ on three species.

\subsection*{Conclusion}%%%%%%%%%%%%%%%%%%%%%%%%%%%%%%%%%%%%%%%%%%%%%%

% Variation of biological objects leads to sampling from moduli spaces.
% whose geometry and combinatorics supply open
% 
% rapidly moving areas of theoretical and applied stratified statistics
% 
% that have geometric and combinatorial or other stratified spaces with
% combinatorial structure.  from these is crucial for statistical
% problems in biology.
% 
% ... a mix of biology, geometry, combinatorics, topology, probability,
% and statistics.
% 
% topics ... tapestry ... interplay

% \comment{here is probably a good place for the very general
% biological phenomics problem, and the potential for mathematics to
% contribute; both the fly wings and the trees projects can be viewed
% from this standpoint}

Spaces of biological forms provide
% a context for
an environment in which mathematical methods
% \comment{referee complaint: spaces of biological forms cannot
% provide mathematical methods.  I disagree, in that once you've got
% your forms in a space, the geometry of the space admits matheamtical
% analysis; maybe rephrase}
can assign distances between phenotypes.
% With such a ``form space'' in hand, ... predict which genes are
% involved given observed phenotype; biologically determine which
% genes are involved, and correlate
The lines of inquiry here fit into biologist David Houle's vision of
phenomics \cite{houle10}, particularly
% his view of
the genotype--phenotype map.  To make a long story short,
% evolution is descent with modification, but
selection acts on phenotype whereas descent and modification act via
genotype, so it becomes desirable to compare phenotypic distance to
genotypic distance,
% needs a metric specifying distance between topologically distinct
% wings
% Also introduce the genotype to phenotype correspondence (cite Houle)
% and its relevance here, which is crucial: if you'll be trying to
including some working definition of each distance in any given case
study.  In general, to grapple with statistical correlations between
genotypes and phenotypes requires a mathematical parametrization for
each of those biological concepts.  For genotypes, that is likely to
involve combinatorial considerations, since the basic quantum of
information is discrete.  Perhaps large-scale continuous analogues or
approximations might be meaningful, as they are for statistical
mechanics,
% in the continuum limit,
% https://www.physicsforums.com/threads/continuum-limit-is-it-really-justifiable.510434
and that is another potential departure point for mathematics.  On the
other hand, phenotype is often continuous in nature: what are the
locations of the vertices in a fly wing?  How do the arcs between the
vertices bow outward or inward?  How do these characteristics change
from wing to wing?  Parametrization in such a context requires
thinking about spaces of continuous objects, the sort of thinking that
mathematics
% excels at.
is designed to carry out.  The examples presented here demonstrate
% the turn in biological thinking when geometric sensibilities are
% taken into account, and similarly
a sample of the kinds of abstract structures in pure
mathematics---along with unexpected questions about them---that
biological investigations reveal.

%%%%%%%%%%%%%%%%%%%%%%%%%%%%%%%%%%%%%%%%%%%%%%%%%%%%%%%%%%%%%%%%%%%%%%%%
%%%%%%%%%%%%%%%%%%%%%%%%%%%%%%%%%%%%%%%%%%%%%%%%%%%
%%%%%%%%%%%%%%%%%%%%%%%%%%%%%%%%%%%%%%%%%%%%%%%%%%%%%%%%%%%%%%%%%%%%%%%%

%%%%%%%%%%%%%%%%%%%%%%%%%%%%%%%%%%%%%%%%%%%%%%%%%%%%%%%%%%%%%%%%%%%%%%%%
\end{document}